\documentclass[apjl]{emulateapj}

\shorttitle{The extended HeII$\lambda$4686-emitting region in IZw18 unveiled}
\shortauthors{Carolina Kehrig et al.}

\usepackage[usenames]{color}

\begin{document}


\title{The extended HeII$\lambda$4686-emitting region in IZw18 unveiled: clues for
  peculiar ionizing sources}

\author{C. Kehrig\altaffilmark{1}}
\author{J. M. V\'\i lchez\altaffilmark{1}}
\author{E. P\'erez-Montero\altaffilmark{1}}
\author{J. Iglesias-P\'aramo\altaffilmark{1,2}}
\author{J. Brinchmann\altaffilmark{3}}
\author{D. Kunth\altaffilmark{4} }
\author{F. Durret\altaffilmark{4} }
\author{F.M. Bayo\altaffilmark{1}}

\affil{\altaffilmark{1}
Instituto de Astrof\'\i sica de Andaluc\'\i a (CSIC), Glorieta de la Astronom\'\i a s/n, E-18008 Granada, Spain.}
\affil{\altaffilmark{2}
Estaci\'on Experimental de Zonas Aridas (CSIC), Ctra. de Sacramento s/n, La Cañada, Almer\'\i a, Spain}
\affil{\altaffilmark{3}
Leiden Observatory, Leiden University, P.O. Box 9513, NL-2300 RA Leiden, The Netherlands}
\affil{\altaffilmark{4}
Institut d'Astrophysique de Paris, UMR 7095, CNRS and UPMC, 98 bis Bd Arago, F-75014 Paris, France}

\begin{abstract}

  New integral field spectroscopy has been obtained for IZw18, the
  nearby lowest-metallicity galaxy considered our best local analog
  of systems forming at high-z. Here we
  report the spatially resolved spectral map of the nebular HeII$\lambda$4686
  emission in IZw18, from which we derived for the first time its
  total HeII-ionizing flux.  Nebular HeII emission implies the
  existence of a hard radiation field. HeII-emitters are observed to
  be more frequent among high-z galaxies than for local objects. So
  investigating the HeII-ionizing source(s) in IZw18 may reveal the
  ionization processes at high-z. HeII emission in star-forming
  galaxies, has been suggested to be mainly associated with Wolf-Rayet stars
  (WRs), but WRs cannot satisfactorily explain the HeII-ionization at
  all times, in particular at lowest metallicities. Shocks from
  supernova remnants, or X-ray binaries, have been proposed as
  additional potential sources of HeII-ionizing photons. Our data indicate that conventional
  HeII-ionizing sources (WRs, shocks, X-ray binaries) are not
  sufficient to explain the observed nebular HeII$\lambda$4686
  emission in IZw18. We find that the HeII-ionizing radiation expected from
  models for either low-metallicity super-massive O stars or rotating
  metal-free stars could account for the HeII-ionization budget
  measured, while only the latter models could explain the highest values of
  HeII$\lambda$4686/H$\beta$ observed. The presence of such peculiar
  stars in IZw18 is suggestive and further investigation in this
  regard is needed. This letter highlights that some of the clues of
  the early Universe can be found here in our cosmic backyard.

\end{abstract}


\section{Introduction}

\label{sec:Intro}

HeII recombination emission indicates the presence of very hard
ionizing radiation with photon energies $\ge$ 54eV. Star-forming
galaxies with lower metallicities tend to have larger nebular
HeII$\lambda$4686 line intensities compared to those with higher
metallicities (e.g. Guseva et al. 2000; Schaerer 2003).  While nebular
HeII emission has been observed in some local low metallicity (Z)
starbursts (e.g. Schaerer et al. 1999; Guseva et al. 2000; Kehrig et
al. 2004; Thuan \& Izotov 2005), HeII-emitters are apparently more
frequent among high-redshift (z) galaxies than for local objects. Recent work has
found that $\ge$ 3$\%$ of the global galaxy population at z$\sim$3
show narrow HeII lines (Cassata et al. 2013), while this number is
much lower at z$\sim$0 (Kehrig et al. 2011). The HeII lines have been
suggested as a good tracer of Population III stars (PopIII-stars; the
first very hot metal-free stars) in high-z galaxies (e.g. Schaerer
2003, 2008). These stars, which should produce a large amount of hard
ionizing radiation, are believed to have contributed significantly to
the Universe's reionization, a challenging subject in contemporary
cosmology (e.g. Bromm 2013). Before interpreting the emission-line
spectra of distant star-forming galaxies, it is crucial to understand
the formation of high-ionization lines in the nearby Universe. The
ideal place to perform this study is in extremely metal-poor nearby
galaxies with nebular HeII emission, which are the natural local counterparts of
distant HeII-emitters.

In this regard, we have been carrying out a programme to investigate
nearby low-Z starburst systems using the integral field
spectroscopy technique (e.g. Kehrig et al. 2008,2013; Perez-Montero et
al. 2011,2013).  As a part of this programme, we have recently
obtained new deep integral field spectroscopic (IFS) data of
IZw18. This is a nearby (D = 18.2 Mpc; Aloisi et al. 2007)\footnote{A
  distance of 18.2 Mpc is assumed in this work} HII galaxy, well known
for its extremely low Z $\sim$ 1/32 solar (e.g. V\'\i lchez \&
Iglesias-P\'aramo 1998), which keeps IZw18 among the three most
metal-poor galaxies known in the local Universe (e.g. Thuan et
al. 2004). Its observational characteristics make IZw18 an excellent
local analogue of primeval systems (see e.g. Lebouteiller et al. 2013
and references therein).

The presence of the nebular HeII$\lambda$4686 line in the spectrum of
IZw18 has been reported before, though the precise location and
extension of this HeII$\lambda$4686 emission is not known
(e.g. Garnett et al. 1991; Izotov et al. 1997; Legrand et al. 1997;
V\'\i lchez \& Iglesias-P\'aramo 1998).  Our unique IFS data unveil for
the first time the entire HeII$\lambda$4686-emitting region and its
structure in IZw18 (see section 3.2).

Despite various attempts to explain it, the origin of the nebular HeII
emission in HII galaxies/regions still remains difficult to understand
in many cases; several potential mechanisms (e.g. hot Wolf-Rayet (WR)
stars, shocks from supernovae remnants, X-ray sources) are proposed
to account (in part or fully) for the HeII ionization in these objects
(e.g. Garnett et al. 1991; Schaerer 1996; Dopita \& Sutherland 1996; Cervi\~no, Mas-Hesse \& Kunth 2002; 
Thuan \& Izotov 2005; Kehrig et al. 2011; Shirazi \& Brinchmann
2012). Though hot WRs have been suggested before as the source of
HeII-ionizing photons in IZw18 (e.g. Izotov et al. 1997; de Mello et
al. 1998), the main mechanism powering the nebular HeII emission in
this galaxy is still an open issue.

In this letter, using new IFS data, we derive for the first time the total
HeII$\lambda$4686-ionizing flux in IZw18 and provide new clues to constrain the
sources of high-ionization.


\section{Integral field spectroscopic data}
\label{sec:Method}

\begin{figure}
\centering
\includegraphics[bb=1 52 1010 903,width=8cm,clip]{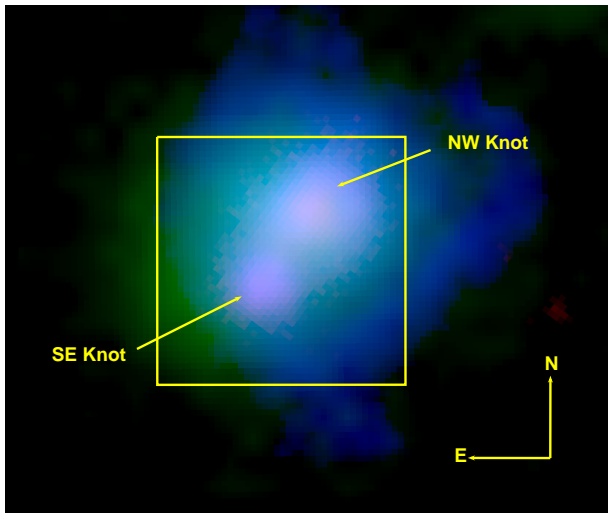}
\caption{Colour-composite image of IZw18 (blue = H$\alpha$ from
  Palomar, green = far-UV/GALEX, red = SDSS r'). The box represents
  the FOV (16''$\times$16'') of the PMAS spectrograph over the galaxy
  main body and the extended H$\alpha$ halo. The PMAS FOV is centered
  on the coordinates RA(J2000.0)= 09$^{h}$:34$^{m}$:02$^{s}$.2  and Dec.(J2000.0)= +55$^{\circ}$:14$^{'}$:25$^{''}$}
\label{pmas_ifu}
\end{figure}

We carried out new IFS observations of IZw18 using the Potsdam
Multi-Aperture Spectrophotometer (PMAS; Roth et al. 2005) on the 3.5 m
telescope at the Calar Alto Observatory (Almeria, Spain). The data
were taken in 2012 December with a typical seeing of 1".  Each spaxel
has a spatial sampling of 1''$\times$1" on the sky resulting in a
field-of-view (FOV) of 16''$\times$16'' ($\sim$ 1.4 kpc $\times$ 1.4
kpc on IZw18; see Figure~\ref{pmas_ifu}). One pointing of IZw18,
encompassing its main body which hosts the two brightest stellar
clusters (referred to as the NW and SE knots), was taken during 2.5 hours integration split into six exposures of 1500 seconds
each.  We used the V500 grating which covers from $\sim$ 3640 to 7200
\AA~ and provides a linear dispersion of $\sim$ 2 \AA/pixel, and full
width at half-maximum effective spectral resolution of $\sim$ 3.6
\AA. Calibration images (exposures of standard star, arc and continuum
lamps) were also obtained. The data reduction was performed following
the procedure described in Kehrig et al. (2013).

\section{Results}

\subsection{Emission-line flux maps}

\begin{figure*}
\centering
\mbox{
  \centering
\includegraphics[bb=50 81 755 709,width=7.5cm,clip]{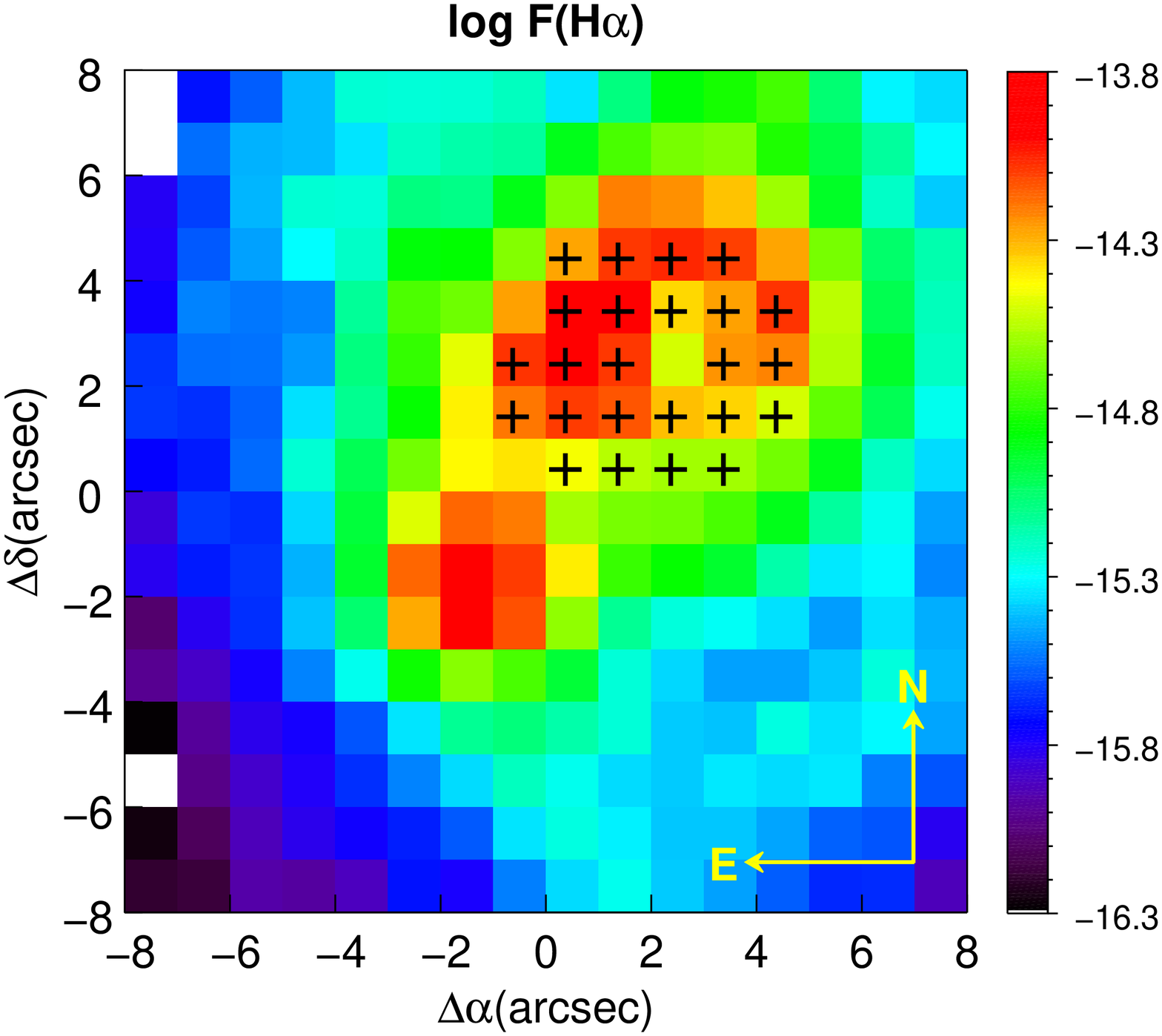} 
\includegraphics[bb=50 81 755 709,width=7.5cm,clip]{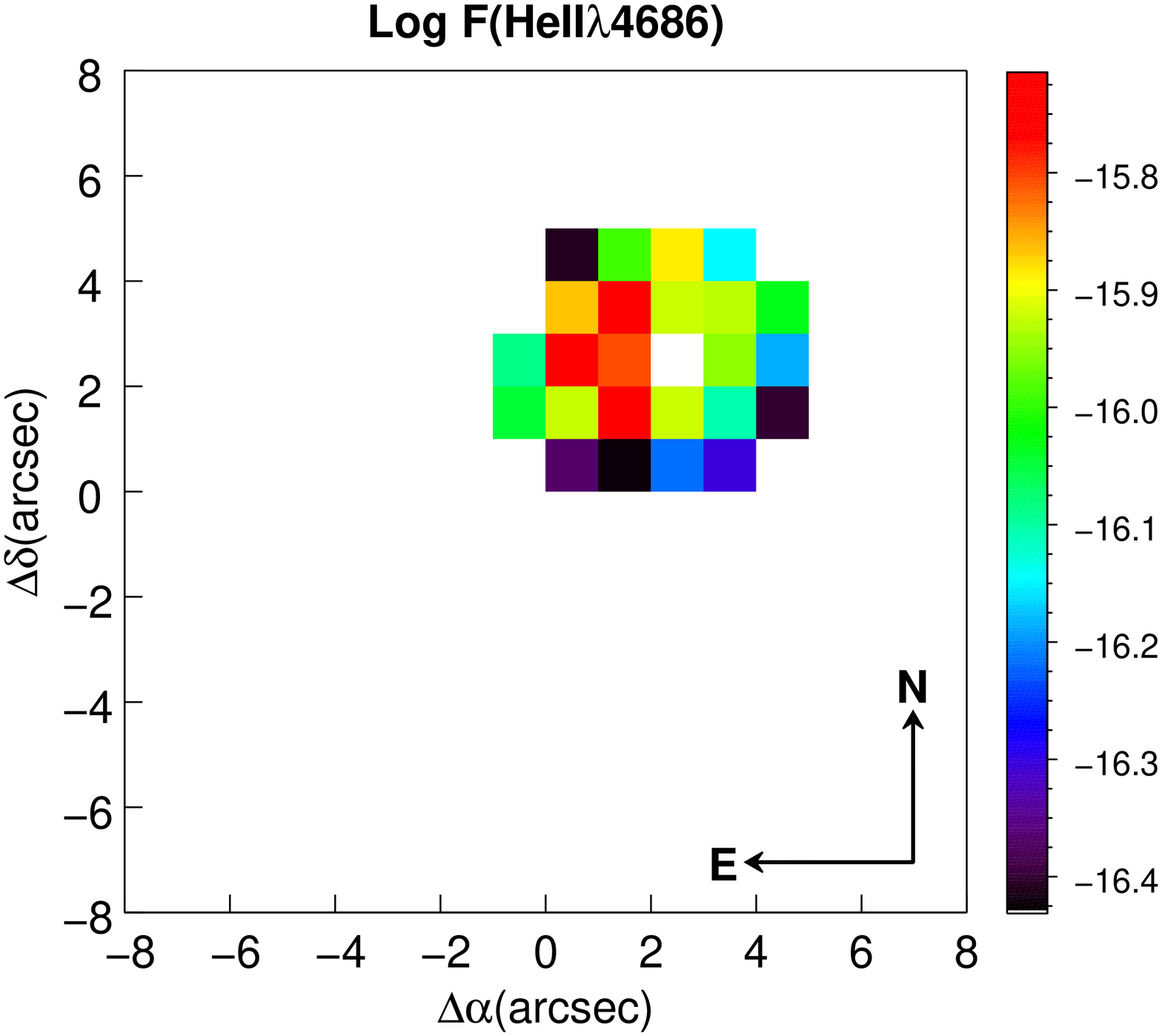}
}
\mbox{
\centering
\includegraphics[bb=50 81 755 709,width=7.5cm,clip]{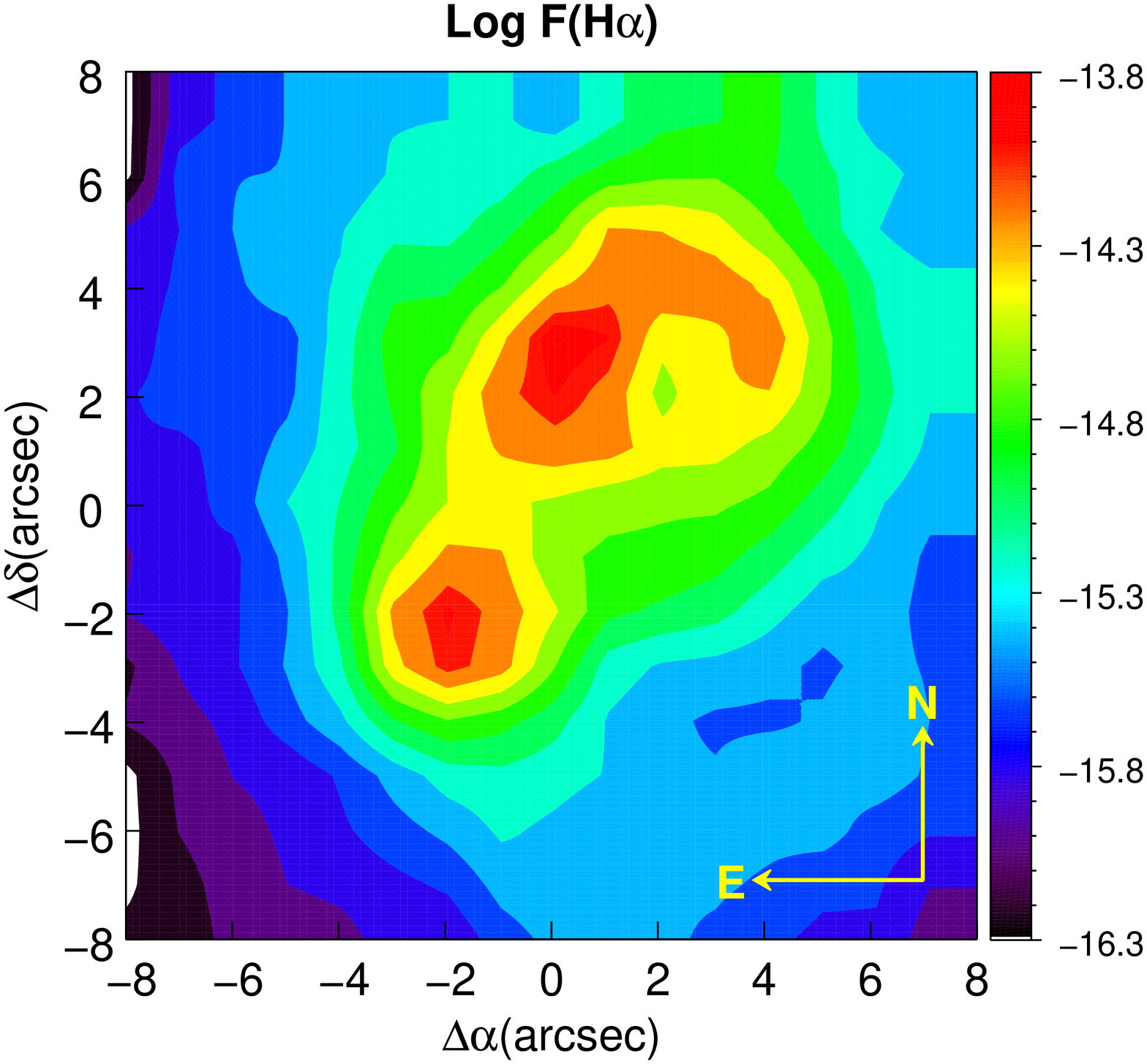} 
\includegraphics[bb=50 81 755 709,width=7.5cm,clip]{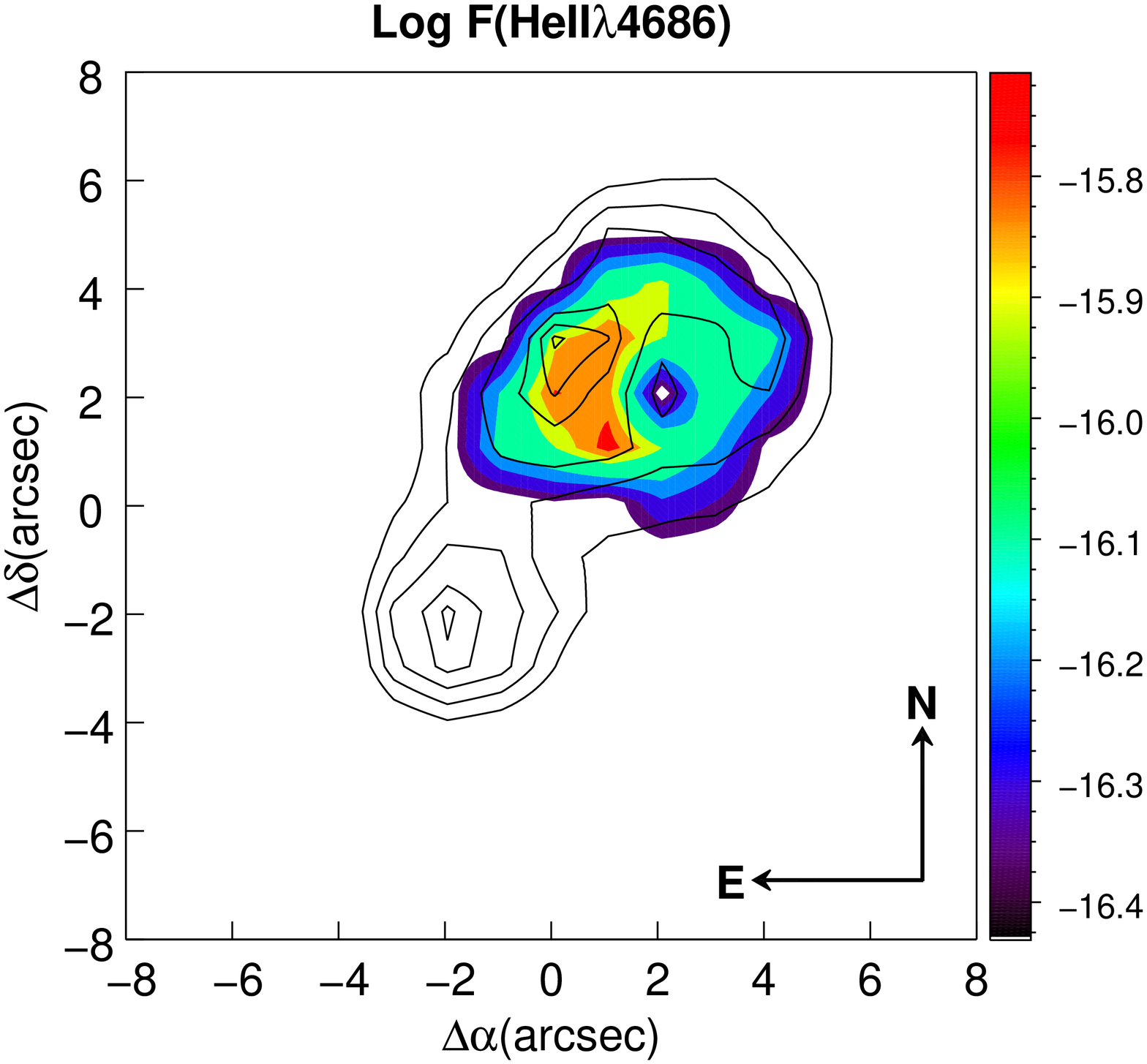} 
}
\mbox{
  \centering
\includegraphics[bb=50 81 755 709,width=7.5cm,clip]{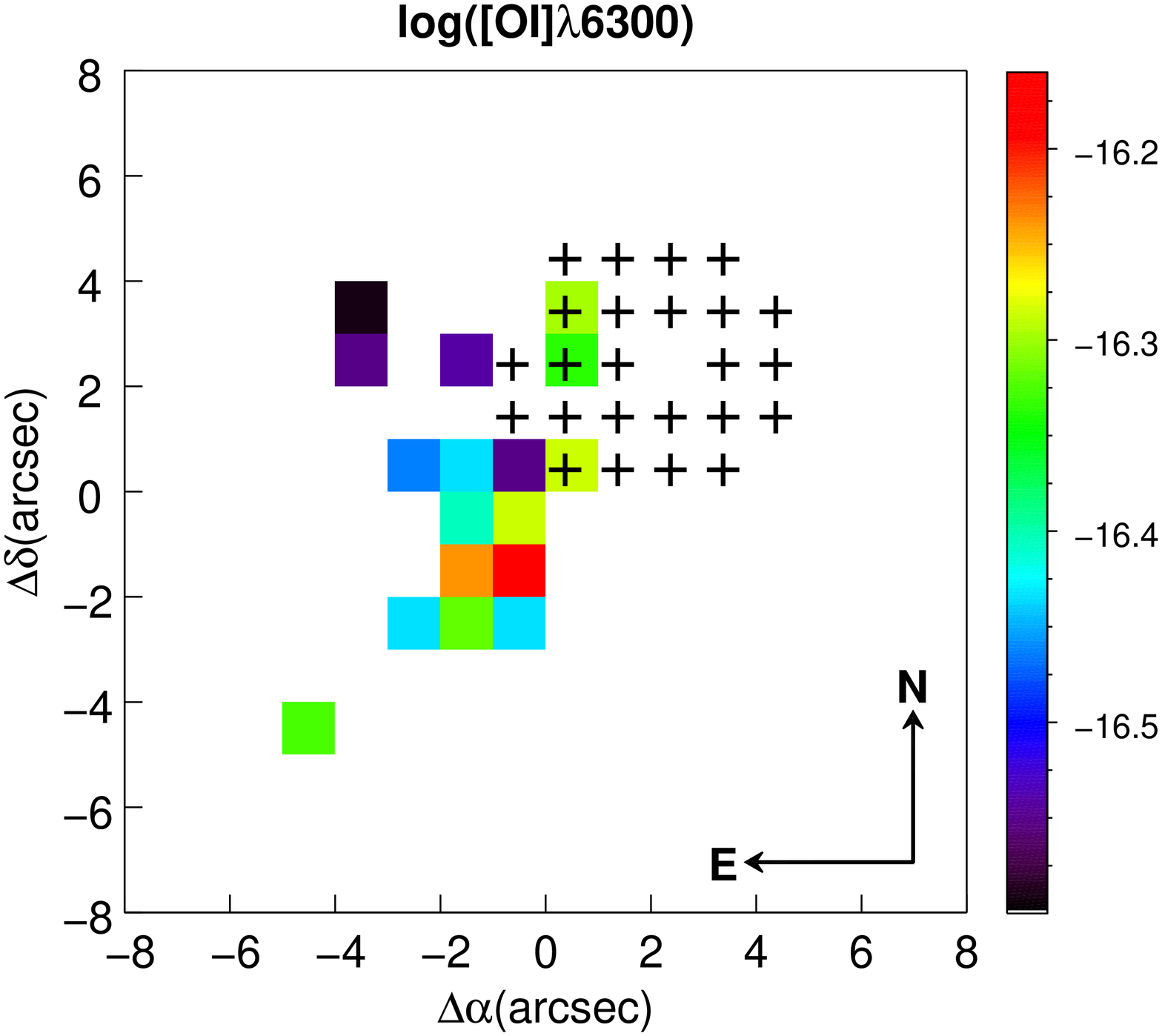} 
\includegraphics[bb=50 81 755 709,width=7.5cm,clip]{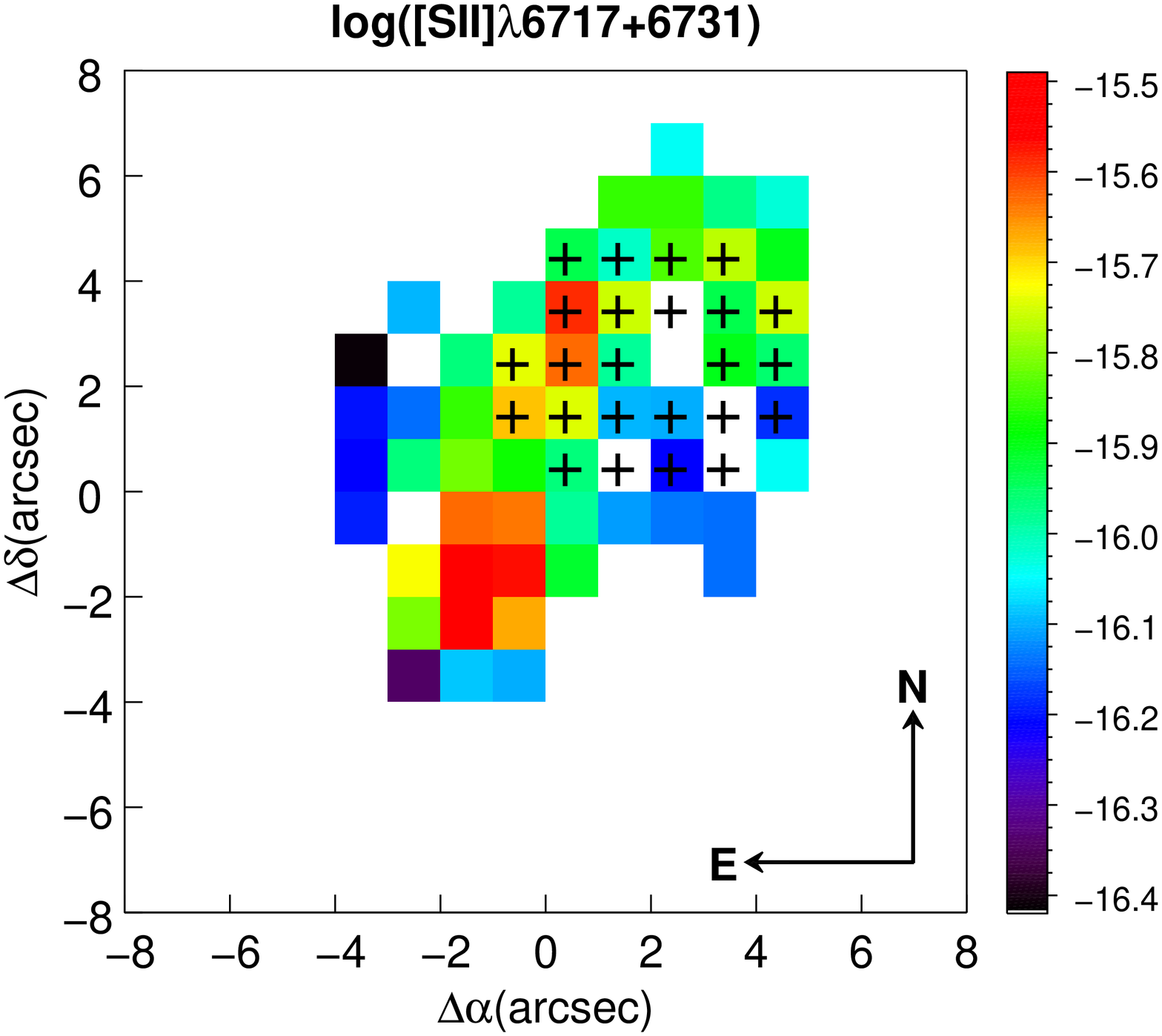} 
}
\caption{Emission-line flux maps of IZw18. Maps are displayed in
  logarithmic scale and the fluxes are in units of erg s$^{-1}$
  cm$^{-2}$; the area of each spaxel is 1 arcsec$^{2}$ on the
  sky. {\it Top row}: H$\alpha$ and
  HeII$\lambda$4686 maps.  {\it Middle row}: for display purposes, the
maps of H$\alpha$ and HeII$\lambda$4686 are presented as color filled
contour plots and were smoothed using bilinear
interpolation. Isocontours of the H$\alpha$ emission line flux are
shown overplotted as reference. {\it Bottom row}: [OI]$\lambda$6300 and [SII]$\lambda$$\lambda$6717+6731 maps. 
The spaxels where we detect
nebular HeII$\lambda$4686 are marked with pluses on the maps of
H$\alpha$ ({\it top row}),  [OI]$\lambda$6300 and [SII]$\lambda$$\lambda$6717+6731 ({\it bottom row}).  The spaxels with
no measurement available are left blank.} 
\label{maps}
\end{figure*}

The emission-line fluxes were measured using the IRAF\footnote{IRAF is distributed by
  the National Optical Astronomical Observatories, which are operated
  by the Association of Universities for Research in Astronomy, Inc.,
  under cooperative agreement with the National Science Foundation.} task SPLOT. The
flux of each emission line was derived by integrating between two
points given by the position of a local continuum placed by eye. For
each line, this procedure was repeated several times by varying the
local continuum position. The final flux of each line and its
associated uncertainty were assumed to be the average and standard
deviation of the independent, repeated measurements (e.g. Kehrig
et al. 2006). 

We used our own IDL scripts to create the emission-line maps presented
in Figure~\ref{maps}. The spaxels where we measure HeII$\lambda$4686
are indicated with pluses.

\subsection{The spatially resolved HeII$\lambda$4686-emitting region}

\begin{figure}
\centering
\includegraphics[bb=18 435 592 700,width=8.5cm,clip]{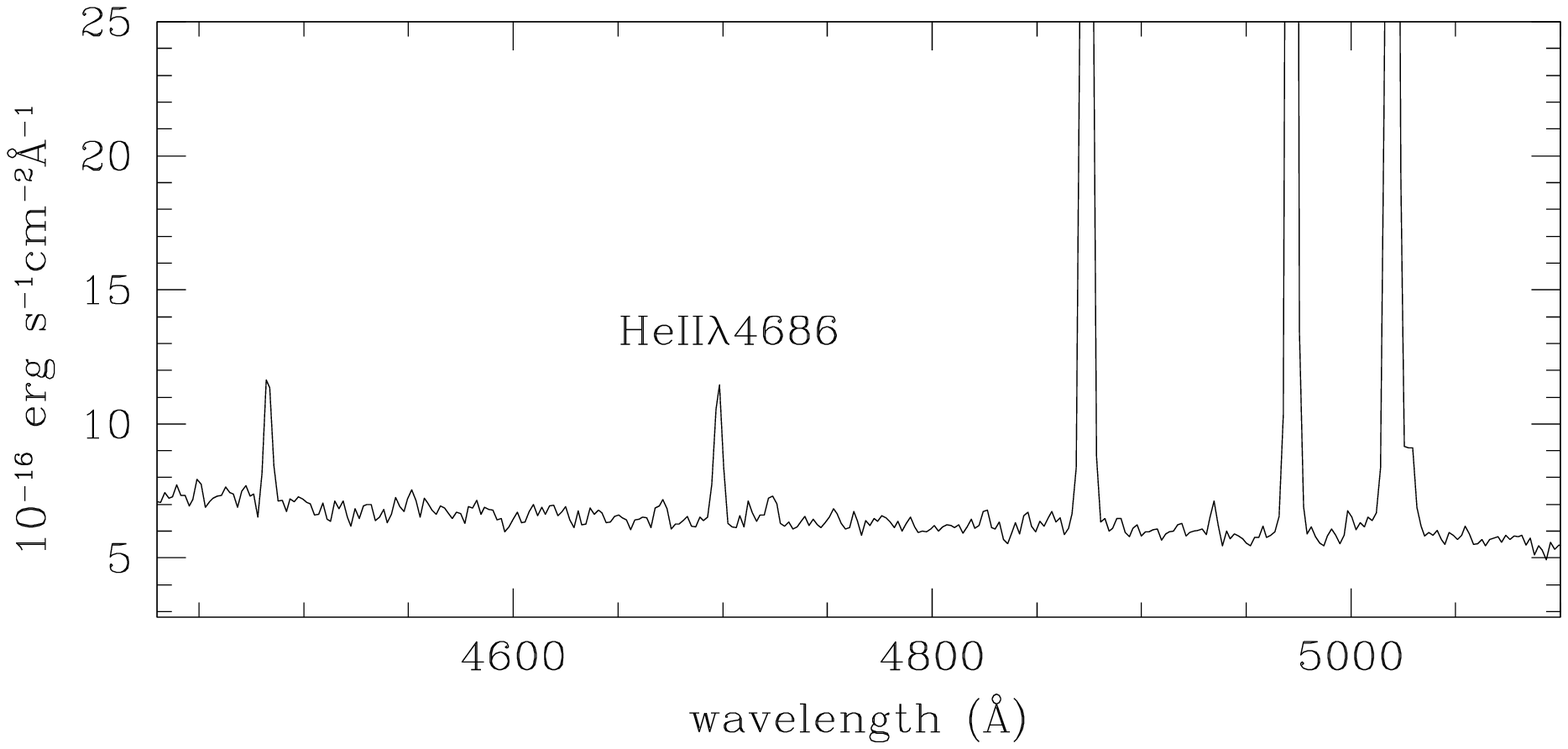}
\caption{Portion ($\sim$ 4425 - 5100 \AA) of the integrated spectrum
  of the HeII$\lambda$4686-emitting region of IZw18.}
\label{spectrum}
\end{figure}

An extended HeII$\lambda$4686-emitting region with a diameter of
$\approx$ 5'' ($\approx$ 440 pc) is revealed from our IFS data (see
Figure~\ref{maps}). The narrow line profile for the HeII$\lambda$4686
emission and its spatial extent are evidence for its nebular nature.
The spectra of certain Of stars may exhibit somewhat narrow
HeII$\lambda$4686 lines (e.g. Massey et al. 2004), so HeII emission
observed in starburst galaxies could be thought to arise in the
atmospheres of such stars (e.g. Bergeron 1977). If these Of stars were
present in appreciable numbers in IZw18, broad H$\alpha$ emission
would however be expected (Massey et al. 2004). Thus, the
non-detection of this feature in our spectra supports the nebular
origin for the HeII$\lambda$4686 line in IZw18.

By adding the emission from the spaxels showing HeII$\lambda$4686 (see
Figure~\ref{maps}), we created the one-dimensional spectrum
for the HeII$\lambda$4686 region (see Figure~\ref{spectrum}). Using
this spectrum, we obtained a very small logarithmic extinction
coefficient c(H$\beta$) = 0.08 $\pm$ 0.02 from the observed ratio
H$\alpha$/H$\beta$ = 2.92 $\pm$ 0.04, assuming an intrinsic case B recombination 
 H$\alpha$/H$\beta$ = 2.75 (Osterbrock \& Ferland 2006, OF06).
 Using PYNEB (Luridiana, Morisset \& Shaw 2015), we derived the
  electron temperature, $T_{e}$ = (2.18 $\pm$ 0.05)$\times$10$^{4}$
    K, and density, $n_{e}$ $\le$ 100 cm$^{-3}$, from the
  [OIII]$\lambda$4363/[OIII]$\lambda$$\lambda$4959,5007 and
  [SII]$\lambda$6717/[SII]$\lambda$6731 ratios, respectively. 
  The integrated flux of the HeII$\lambda$4686 line,
corrected for reddening, is (2.84 $\pm$ 0.18)$\times$10$^{-15}$ erg
s$^{-1}$ cm$^{-2}$ which translates to a HeII$\lambda$4686 luminosity
of L$_{HeII\lambda4686}$ = (1.12 $\pm$ 0.07)$\times$10$^{38}$ erg
s$^{-1}$. The corresponding HeII ionizing photon flux, Q(HeII)$_{obs}$
= (1.33 $\pm$ 0.08)$\times$10$^{50}$ photon s$^{-1}$, was derived from
the measured L$_{HeII\lambda4686}$ using the relation Q(HeII) =
L$_{HeII\lambda4686}$/[j($\lambda$4686)/$\alpha_{B}$(HeII)] (assuming
case B recombination, and $T_{e}$ = 2$\times$10$^{4}$ K; OF06).  
We checked that different CLOUDY models (Ferland et al. 2013) computed 
for the typical $n_{e}$ and $T_{e}$ in the HeII$\lambda$4686 region, and
considering several effective temperatures and geometries with no
dust, give a ratio L(HeII$\lambda$4686)/Q(HeII) which agrees with the
assumed ratio (OF06) within 10$\%$. The total
Q(HeII)$_{obs}$, a quantity not reported before for IZw18, will allow
us to constrain possible ionizing sources of HeII in IZw18.

\section{Discussion}

\subsection{Ionizing sources of HeII}

One widely favored mechanism for HeII ionization in HII galaxies
involves hot WRs (e.g. Schaerer 1996). Nevertheless, it has been
demonstrated that nebular HeII$\lambda$4686 does not appear to be
always associated with WRs, as it is the case of the HeII nebulae
  LMC-N44C, LMC-N159F and M33-BCLMP651, among others (Kehrig et al. 2011 and
  references therein). This indicates that WRs cannot explain HeII
  ionization in all cases, particularly at low-Z
  (e.g. Guseva et al. 2000; Kehrig et al. 2008,2013; Schaerer 2003;
  Shirazi \& Brinchmann 2012). Besides, it has been shown that, for
a sample of HeII$\lambda$4686-emitting star-forming galaxies, current
models of massive stars predict HeII$\lambda$4686 emission, and
HeII$\lambda$4686/H$\beta$ ratios, only for Z $>$ 0.20 Z$_{\odot}$
instantaneous bursts (Shirazi \& Brinchmann 2012).  In the particular
case of IZw18, previous work has claimed that the ratio
HeII$\lambda$4686/H$\beta$ could be reproduced using highly
density-bounded photoionization models while underpredicting the
electron temperature measurement (Stasi{\'n}ska \& Schaerer 1999);
these models have been challenged by P{\'e}quignot (2008).

Faint broad emission signatures, attributable to WRs, are observed
in the spectrum of IZw18 despite its low-Z (e.g. Izotov et al. 1997, Legrand
et al. 1997, Brown et al. 2002). A comprehensive study of WRs in
IZw18, using UV STIS spectroscopy, revealed signatures of carbon-type
WRs (WC) in two clusters: one in the NW star-forming region and a
second one on the outskirts of this region (Brown et al. 2002). Here
we deal with the NW one, since it is in the HeII$\lambda$4686-emitting
region (see Figure~\ref{maps}). The CIV$\lambda$1550 flux measured
from the NW WR cluster (Brown et al. 2002) gives a CIV$\lambda$1550
luminosity of L$_{1550}$ = 4.67$\times$10$^{37}$ erg s$^{-1}$. Taking
the L$_{1550}$ luminosity of the metal-poor early-type WC (WCE) model
by Crowther \& Hadfield (2006, CH06), which mimics a single WCE star in
IZw18, this implies $\sim$ 9 IZw18-like WCE stars present in the NW
cluster. From these 9 WCE stars a total flux Q(HeII) =2.8
$\times$10$^{48}$ photon s$^{-1}$ is expected [assuming Q(HeII) =
  10$^{47.5}$ photon s$^{-1}$ for one IZw18-like WCE; CH06], i.e. about 48 times lower than the
Q(HeII)$_{obs}$=(1.33$\pm$0.08)$\times$10$^{50}$ photon s$^{-1}$,
derived from our data (see section 3.2).

Based on the HeII-ionizing flux expected from these IZw18-like WRs, a
  very large WR population is required to explain the HeII-ionization
  budget measured; for instance, taking the Q(HeII)= 10$^{47.5}$
  photon s$^{-1}$ for one IZw18-like WCE (CH06),
  the number of these WCEs needed to explain our derived Q(HeII)$_{obs}$ =
  (1.33$\pm$0.08)$\times$10$^{50}$ photon s$^{-1}$ would be $>$ 400.
In principle, the presence of hundreds of WRs in IZw18 should not be
discarded on the basis of empirical arguments for reduced WR line
luminosity at low-Z (CH06).  However,
  assuming a Salpeter initial mass function (IMF; Salpeter 1955;
  M$_{up}$=150M$_\odot$) and the initial mass needed for a star to
  certainly become a WC (Maeder \& Maynet 2005), a cluster with $>$ 8
  times the total stellar mass of the NW region (M$_{*,NW}$ =
  2.9$\times$10$^{5}$ M$_{\odot}$ from Stasi\'nska \& Schaerer (1999)
  scaled to 18.2 Mpc distance) is
  required to provide $>$ 400 WC in IZw18. Also, such a high number
of WRs is not supported by state-of-the-art stellar evolutionary
models for single (rotating and non-rotating) massive stars in 
metal-poor environments (Leitherer et al. 2014).  Furthermore,
  given the decrease in the ratio of WR/O stars with decreasing metallicity,
  shown by observations and theoretical models (Maeder \& Meynet 2012
  and references therein), such a large number of WRs appears clearly
  unreasonable considering the extremely low-Z  and the O star content of IZw18
  (CH06). All this suggests that WRs are not
the sole responsible of the HeII$\lambda$4686 emission in IZw18.

The binary channel in massive star evolution is suggested to increase
the WR population (e.g. Eldridge et al. 2008), but the WR
population in Local Group galaxies does not show an increased binary
rate at lower-Z (e.g. Foellmi et al. 2003;
Neugent \& Massey 2014); thus the binary channel does not seem to
favour the formation of WRs at lower-Z, in contrast to what we
need. Nevertheless, we should bear in mind the possible uncertainties yet
unsolved in the models (Maeder \& Meynet 2012). Further
investigation awaits the calculation of evolutionary models for binary
stars at very low-Z.

As mentioned before, nebular HeII$\lambda$4686 emission observed in
the spectra of HII galaxies and extragalactic HII regions has often
been attributed also to shocks and X-ray sources (e.g. Pakull \&
Mirioni 2002; Garnett et al. 1991; Thuan \& Izotov 2005). In the
following we discuss on these two candidate sources for HeII
ionization in IZw18. 

The X-ray emission from IZw18 is dominated by a single X-ray binary
apparently located in the field of the NW knot (Thuan et al. 2004). We
have computed a CLOUDY photoionization model (Ferland et al. 2013)
using as input a power-law spectral energy distribution (SED) with the
same X-ray luminosity, column density and slope that have been
reported for IZw18 (Thuan et al. 2004). This CLOUDY model gives as an
output a HeII$\lambda$4686 luminosity L$_{HeII\lambda4686}$ =
10$^{35.7}$ erg s$^{-1}$ which is $\sim$ 100 times lower than the
L$_{HeII\lambda4686}$ measured. This result rules out the X-ray binary
as the main source of HeII ionizing photons in IZw18.  We note here
that the emission from X-ray ionized nebulae has been successfully
reproduced by CLOUDY models before (e.g. Pakull \& Mirioni 2002).

Guided by the existence of HeII$\lambda$4686 emission associated with
supernova remnants (e.g. Kehrig et al. 2011) we explored the
conjecture that the HeII$\lambda$4686 region represents such a
shock-ionized nebula. The [OI]$\lambda$6300 line, often strong in
remnants, has been frequently used as a sensitive shock-emission test
(e.g. Skillman 1985).  We checked that, in fact, most of
the [OI]$\lambda$6300 emission in IZw18 is concentrated on the SE knot
(see the [OI]$\lambda$6300 map in Figure~\ref{maps}),
with only 12$\%$ of the HeII$\lambda$4686-emitting spectra showing
[OI]$\lambda$6300 flux above the 3$\sigma$ detection
limit. Additionally, we find no evidence for [SII] enhancement (a
usual sign of shock excitation; e.g. Dopita \& Sutherland 1996)
associated to the location of the
HeII$\lambda$4686 region (see the [SII] map in Figure~\ref{maps}). Therefore, the
HeII$\lambda$4686-emitting zone in IZw18 is unlikely to be produced by shocks.

\subsection{Peculiar very hot stars in IZw18 ?}

Our new observations have allowed us to empirically demonstrate why
conventional HeII-ionizing sources (e.g. WRs, shocks, X-ray
binaries) cannot account for the total HeII-ionization budget in
IZw18. What could the nebular HeII$\lambda$4686 emission in IZw18
originate from ?

We have also explored the possibilty of very massive, metal-poor O stars to
account for our observations of IZw18. Using current wind models of
very massive O stars at low-Z we can derive their HeII-ionizing fluxes
(Kudritzki 2002, K02). According to the hottest models ($T_{e}$=60,000 K),
between 10-20 super-massive stars with 300M$_{\odot}$ [with Q(HeII)
  $\approx$ (0.70-1.4)$\times$10$^{49}$ photon s$^{-1}$ each] would be
enough to explain the derived Q(HeII)$_{obs}$ budget. Very massive
stars of up to 300M$_{\odot}$ were claimed to exist in the LMC-R136
cluster (Crowther et al. 2010); however the existence of such
super-massive 300M$_{\odot}$ stars remains heavily debated  (Vink 2014). Additionally, assuming a Salpeter IMF,
  10-20 stars with 290 $\le$ M$_{*}$/M$_{\odot}$ $\le$ 310  would imply a cluster
  mass of $\sim$ 10-20 $\times$ M$_{*,NW}$.  We should bear in mind that an extrapolation of the IMF
  predicting 300M$_{\odot}$ stars remains unchecked up to now. If we consider instead
the 150M$_{\odot}$ star hottest models [with Q(HeII) $\le$
  1.9$\times$10$^{47}$ photons$^{-1}$ each, for Z $\le$ 1/32 Z$_{\odot}$;
  K02], the number of these stars required to explain the
Q(HeII)$_{obs}$ would be $>$ 650. For a Salpeter IMF, 650 stars
with 145 $\le$ M$_{*}$/M$_{\odot}$ $\le$ 155 would require a cluster mass $\sim$ 200 $\times$
M$_{*,NW}$. Besides the Q(HeII)$_{obs}$ budget, in the
HeII$\lambda$4686 region we have measured HeII$\lambda$4686/H$\beta$
ratios as high as 0.08. These values appear too big to be explained
even by the models for the hottest, most metal-poor super-massive
300M$_{\odot}$ stars (K02) under ionization-bounded
conditions.  Further constraints to the observations should await the
calculation of new evolutionary tracks and SEDs for single O stars at
the metallicity of IZw18 including rotation.

Searches for very metal-poor starbursts and PopIII-hosting galaxies
have been carried out in the distant Universe using the HeII lines
(e.g. Schaerer 2008; Cassata et al. 2013).
This search is based on the high effective temperature for
PopIII-stars which will emit a large number of photons with energy above 54eV, and
also on the expected increase of the HeII recombination lines with
decreasing Z (e.g. Guseva et al. 2000; Schaerer 2003,2008).
Predictions for burst models of different metallicities show how
their corresponding Q(HeII) can increase by up to $\sim$ 10$^{3}$ when
going from Z=10$^{-5}$ to 0 (Schaerer 2003). These models cannot
explain the Q(HeII)$_{obs}$ when Z $\ge$ 10$^{-5}$ [for a Salpeter
IMF, M$_{up}$ = 100M$_{\odot}$; see table 3 in Schaerer 2003] even
assuming that the total M$_{*,NW}$ would
come from HeII-ionizing stars. So another more speculative possibility, to explain the derived
Q(HeII)$_{obs}$ could be based on nearly metal-free ionizing stars. These stars should ionize HeII via their strong UV
  radiation expected at nearly zero metallicity (e.g. Tumlinson \& Shull 2000;
  Schaerer 2003). 

As an approximation of nearly metal-free single stars, we have 
compared our observations with the HeII-ionizing radiation expected
from state-of-the-art models for rotating Z=0 stars (Yoon, Dierks \&
Langer 2012). According to these models, we found that a handful of
such stars could explain our derived Q(HeII)$_{obs}$ [e.g. $\sim$ 8-10
  stars with mass M$_{ini}$=150M$_{\odot}$ or $\sim$ 13-15 stars with
  M$_{ini}$=100M$_{\odot}$; with Q(HeII) $\approx$
  1.4$\times$10$^{49}$ (0.9$\times$10$^{49}$) photon s$^{-1}$ for each
  star with M$_{ini}$ =150M$_{\odot}$(100M$_{\odot}$)]. Additionally,
we note that the ionizing spectra produced by these star models are
harder than the ones expected from the hottest models of super-massive
300M$_{\odot}$ stars (K02), so they would also explain the
highest HeII$\lambda$4686/H$\beta$ values observed, providing that
ionization-bounded conditions are met. While gas in IZw18 is
very metal-deficient but not primordial, Lebouteiller et al. (2013) have
pointed out that the HI envelope of IZw18 near the NW knot contains
essentially metal-free gas pockets.  These gas pockets
could provide the raw material for making such nearly metal-free 
stars. Clearly, in this hypothetical scenario, these extremely metal-poor stars cannot 
belong to the NW cluster, which hosts more chemically evolved stars.

\section{Summary and concluding remarks}

This letter reports on new optical IFS observations of the nearby dwarf
galaxy IZw18. This is an extremely metal-poor system, which is our
best local laboratory for probing the conditions dominating in distant 
low-Z starbursts.  Our IFS data reveal for the first time
the total spatial extent ($\approx$ 440 pc diameter) of the
HeII$\lambda$4686-emitting region and corresponding total HeII-ionizing photon
flux in IZw18.  The metal-poor sensitivity of the HeII line is a primary motivation to develop
diagnostics for unevolved starbursts, and strong nebular
HeII emission is expected to be one of the best signatures of massive
PopIII-stars (e.g. Schaerer 2003,2008). HeII emission has been observed to
be more frequent at higher-z than locally (Kehrig et
 al. 2011; Cassata et al. 2013). Thus the analysis of the origin of the
HeII$\lambda$4686 nebular line in relatively close ionized regions,
which can be studied in more detail, can yield insight into the
ionizing sources in the distant Universe. 

Our observations combined with stellar models predictions point out that conventional
excitation sources (e.g. WRs, shocks, X-ray binaries) cannot
convincingly explain the HeII-ionizing energy budget derived for
IZw18. Other mechanisms are probably also at work. If the
HeII-ionization in IZw18 is due to stellar sources, these might be
peculiar very hot stars (perhaps uncommon in local starbursts but
somewhat more frequent in distant galaxies): according to theoretical stellar
models, either super-massive O stars or nearly metal-free ionizing
stars could in principle account for the total Q(HeII)$_{obs}$ of IZw18. However the
super-massive O stars scenario would imply a cluster mass much 
higher than the mass of the NW knot derived from observations. On the other hand, 
though metal-free gas pockets  were previously reported in IZw18 (Lebouteiller et al. 2013), we 
highlight that the existence of nearly metal-free ionizing stars is not yet confirmed observationally.
The work presented here can help in the preparation of prospective searches for primeval
objects, one of the main science drivers for next-generation telescopes (e.g. Bromm 2013).


\begin{acknowledgements}

This work has been partially funded by research projects
AYA2010-21887-C04-01 from the Spanish PNAYA, and PEX2011-FQM7058 from
Junta de Andalucia. FD and DK gratefully acknowledge support from the
Centre National d'Etudes Spatiales. We thank the anonymous referee for
a constructive report which improved this letter.  We express our appreciation to
Leslie Sage for his help and suggestions. We also would like to thank
Manfred Pakull for his useful comments on this letter.

\end{acknowledgements}

\end{document}